\documentclass[conference]{IEEEtran}
\IEEEoverridecommandlockouts
\usepackage{cite}
\usepackage{amsmath,amssymb,amsfonts}
\usepackage{algorithmic}
\usepackage{graphicx}
\usepackage{textcomp}
\usepackage{booktabs}
\usepackage{orcidlink}
\def\BibTeX{{\rm B\kern-.05em{\sc i\kern-.025em b}\kern-.08em
    T\kern-.1667em\lower.7ex\hbox{E}\kern-.125emX}}

\begin{document}

\title{Toward a Quantum Information System Cybersecurity Taxonomy and Testbed: Exploiting a Unique Opportunity for Early Impact \thanks{The work presented in this paper was partially supported by the U.S. Department of Energy, Office of Science under DOE contract number DE-AC02-06CH11357. The submitted manuscript has been created by UChicago Argonne, LLC, operator of Argonne National Laboratory. Argonne, a DOE Office of Science laboratory, is operated under Contract No. DE-AC02-06CH11357. The U.S. Government retains for itself, and others acting on its behalf, a paid-up nonexclusive, irrevocable worldwide license in said article to reproduce, prepare derivative works, distribute copies to the public, and perform publicly and display publicly, by or on behalf of the Government.}}

\author{
\IEEEauthorblockN{Benjamin Blakely\orcidlink{0000-0002-2533-618X}, Ryan Syed, Alec Poczatek}
\IEEEauthorblockA{\textit{Strategic Security Sciences} \\
\textit{Argonne National Laboratory}\\
bblakely@anl.gov, rsyed@anl.gov, poczatek@anl.gov}
\and
\IEEEauthorblockN{Joaquin Chung\orcidlink{0000-0001-7383-3810}, Raj Kettimuthu}
\IEEEauthorblockA{\textit{Data Science and Learning} \\
\textit{Argonne National Laboratory}\\
chungmiranda@anl.gov, kettimut@anl.gov}
}

\maketitle

\begin{abstract}
Any human-designed system can potentially be exploited in ways that its designers did not envision, and information systems or networks using quantum components do not escape this reality. We are presented with a unique but quickly waning opportunity to bring cybersecurity concerns to the forefront for quantum information systems before they become widely deployed. The resources and knowledge required to do so, however, may not be common in the cybersecurity community. Yet, a nexus exist. Cybersecurity starts with risk, and there are good taxonomies for security vulnerabilities and impacts in classical systems. In this paper, we propose a preliminary taxonomy for quantum cybersecurity vulnerabilities that accounts for the latest advances in quantum information systems, and must evolve to incorporate well-established cybersecurity principles and methodologies. We envision a testbed environment designed and instrumented with the specific purpose of enabling a broad collaborative community of cybersecurity and quantum information system experts to conduct experimental evaluation of software and hardware security including both physical and virtual quantum components. Furthermore, we envision that such a resource may be available as a user facility to the open science research community.
\end{abstract}


\section{Introduction}

Quantum information sciences (QIS) is a broad field spanning multiple scientific communities. At the most fundamental level, it is a set of problems for physicists, though the overlaps with computer science and information theory are difficult to fully disentangle. While cybersecurity as a science has been discussed in classical systems, it has remained elusive~\cite{siponen_critical_2007, schneider_blueprint_2012, kott_science_2015, chang_synergy_2016, herley_sok_2017, burton-jones_examining_2021}. As applications of quantum-based systems and networks become increasingly practical, communities such as network operators and cybersecurity professionals become more involved. Quantum computing has gotten considerable press, and in fact has seen much progress over the past few years~\cite{gill_quantum_2022}. Quantum supremacy, the demonstration that a quantum computer can in-fact out perform a classical computer, has already been realized~\cite{arute_quantum_2019}.

Of nearly equal interest have been the implications of communication networks based on quantum principles. The road to a fully ``quantum internet'' might be long, but it has already been envisioned\cite{wehner_quantum_2018}. Early protocols to enable such a vision have seen successes, but also challenges\cite{kozlowski_designing_2020}. Determining how to extend the range of quantum communications beyond the limits of optical attenuation, in a manner that does not compromise the security and throughput that drive interest in these technologies in the first place, is an important research problem. 

A less commonly identified area of QIS is that of quantum sensing. Quantum sensors allow the measurement of physical properties with unprecedented sensitivity and precision\cite{degen_quantum_2017}. Quantum sensors, as an input to quantum computers and networks, will thus be a critical technology to fully realizing the power of quantum computing as it pertains to interactions with the real world. 

In order to leverage these technologies, it is critical that security assurances can be provided. Given the fundamentally different nature of QIS, new protocols are required (or enabled). For example, an entirely new network stack (the combination of software components translating digital information into their physical manifestation and back) might be required\cite{dahlberg_link_2019,qi-stack-survey}.

Entanglement is the basis for much of quantum communication security, but this only works when there is a direct channel to exchange entangled particles. Optical communications provide high bandwidth to exchange entangled photons, but attenuation losses limit their range. Constructing point to point links between all pairs of communicating nodes will be infeasible. Thus, an important area of research is how to construct repeaters to distribute or extend quantum communications beyond a single link without compromising the security guarantees\cite[e.g.]{li_long_2013, jones_design_2016}.

One of the most-studied methods of leveraging quantum channels for secure communications is Quantum Key Distribution (QKD) algorithms. In these, strong classical symmetric encryption algorithms such as AES~256 are used, but a quantum channel is used to agree upon a session key for each communication. This provides provably perfect secrecy of the key because any attempt to eavesdrop will be immediately detectable at a physical level\cite{nandal_survey_2021}. However, this is a complex scheme to implement and may decrease efficiency \cite{jasim_quantum_2015}.

The resilience to tampering that enables QKD protocols can also be used in other authentication contexts -- those where assurances against tampering and forgery are needed -- and many studies have been conducted in this area\cite{majumdar_sok_2021}. However, researchers have found weaknesses related to forge-ability\cite{doosti_unified_2021} and clone-ability\cite{majenz_limitations_2021} in some algorithms, showing the importance of understanding nuances in implementations of quantum algorithms. In classical encryption schemes, it has often been the implementation and not the algorithm that creates a vulnerability, and care must be taken not to repeat these mistakes in QIS.

Protocols have also been proposed for quantum metrology\cite{eldredge_optimal_2018} (i.e., measurements using quantum sensors). These protocols must also provide assurances of their security if they are to be used in sensitive application domains. The nature of quantum information once again makes tampering with measured quantities impossible without potential detection, but this sensitivity also can create additional susceptibility to noise.

\section{A Taxonomy for Quantum Cybersecurity}

All QIS developments are understandably exciting from the perspective of new capabilities they will enable -- whether it be algorithmic performance, new insight into the natural world, or seemingly iron-clad guarantees about communications security. Still, the history of the classical Internet is one filled with sand traps and land mines. While considerable work on specific security considerations for QIS-based systems have been conducted, we have yet to find a comprehensive taxonomy of the specific threat vectors applicable to this domain. It is possible to apply more general works, but this might fail to properly calibrate attention on the types of threats that are only feasible, are infeasible, or are fundamentally different in a quantum versus classical domain. 

We thus suggest that a critical next-step in developing a comprehensive discipline of security engineering and academic study in the QIS domain is creating such a taxonomy. Partnering cybersecurity experts with physicists and computer scientists is the only way to provide assurance as to the comprehensiveness of such work. In the discussion that follows, we provide a sketch of taxonomy based on the MITRE Common Attack Pattern Enumeration and Classification (CAPEC) framework\cite{noauthor_capec_2021} and Common Weakness Enumeration (CWE) framework\cite{noauthor_cwe_2019}. 

The CWE framework includes a large number of hardware-related considerations, broken up into the following categories (with indicated sub-item counts):

\begin{itemize}
    \item Manufacturing and Life Cycle Management Concerns - (1195)
    \item Security Flow Issues - (1196)
    \item Integration Issues - (1197)
    \item Privilege Separation and Access Control Issues - (1198)
    \item General Circuit and Logic Design Concerns - (1199)
    \item Core and Compute Issues - (1201)
    \item Memory and Storage Issues - (1202)
    \item Peripherals, On-chip Fabric, and Interface/IO Problems - (1203)
    \item Security Primitives and Cryptography Issues - (1205)
    \item Power, Clock, and Reset Concerns - (1206)
    \item Debug and Test Problems - (1207)
    \item Cross-Cutting Problems - (1208)
    \item Physical Access Issues and Concerns - (1388)
\end{itemize}

Of these, some do not seem likely to differ greatly from classic to quantum contexts: Manufacturing and Life Cycle Management Concerns, Security Flow Issues (the way control is handed off from component to component within a system), Integration Issues, Debug and Test Problems, and Cross-cutting Concerns. Others have potential specific quantum nuances. 

Privilege Separation and Access Control Issues may become much different if multi-user quantum computers become commonplace. For example, as the operation of a quantum processor is dependent upon a coherent state, is it possible to have a concurrent multi-user systems in this context? How does one partition the quantum computations within these processors to be sufficiently isolated while still having robust coherence? Schemes such as Time Division Multiple Access (TDMA) for quantum communication channels or buffer zones for quantum processors give potential options, but come at a cost of efficiency.

The category of Security Primitives and Cryptography issues is one area that has received significant attention to-date. Post-quantum ciphers are being developed to resist the threat of quantum computing attacks. However, these ciphers have occasionally proven to be vitally flawed in themselves (e.g., \cite{castryck_efficient_2022}). As of July 2022, NIST has approved one algorithm for general encryption, and three for digital signatures\footnote{\url{https://csrc.nist.gov/Projects/post-quantum-cryptography/selected-algorithms-2022}}. This will no doubt be an area of continuing research, especially given that three of the four currently approved algorithms are reliant on the same family of math problems.

The manner in which circuit and logic devices are secured will be fundamentally different in many cases due to the difference in foundational technologies. Ensuring security with photonic systems, for example, may be quite different than electrical components. This also impacts Core and Compute issues as new instruction sets are required for quantum processors to leverage these new components and coordinate quantum computations. Preventing the occurrence of SPECTRE or Meltdown-type flaws\footnote{https://meltdownattack.com/} in hardware design may not be a priority unless the issue is highlighted in the QIS context. Likewise, memory and storage issues must be considered to prevent future ROWHAMMER-type attacks \cite{mutlu_rowhammer_2017}. Similar concerns are relevant to the categories of Power, Clock, and Reset Concerns; Peripherals, On-chip Fabric, and Interface I/O Problems; and Physical Access Issues and Concerns -- i.e., are there ways to cause quantum processors or sensors to perform in unexpected ways by tampering with their physical components, inputs, state, or control signals?

CWE's focus on hardware flaws is a very helpful approach to start from in building a quantum security taxonomy. It does miss some elements, however, that are more focused on the attack vector versus the flaw itself. For example, interception of communications is a flaw either in a hardware device or software algorithm, but doesn't clearly fit into a CWE category. This is where the CAPEC framework can help by adding another dimension to our considerations. As there are over 800 items in the CAPEC taxonomy, and even at the level of meta- or standard-attack patterns a full enumeration is too unwieldy for this paper, we will highlight a number of meta-attack patterns within their category that are of particular interest. For our purposes, we will consider the categories of Supply Chain, Social Engineering, and Physical Security to be largely overlapping with classical security considerations and exclude them here. This leaves the categories of software-, hardware-, and communication-based attack patterns. The list of categories and their sub-item count is as follows:

\begin{itemize}
    \item Software - (513)
    \item Hardware - (515)
    \item Communications - (512)
    \item Supply Chain - (437)
    \item Social Engineering - (403)
    \item Physical Security - (514)
\end{itemize}

Software meta-patterns related to the interaction between running code and underlying hardware as it pertains to quantum state (as above in the Privilege Separation and Access Control Issues discussion) are particularly relevant in that category. Of the several dozen meta-patterns, the following may be worth quantum-specific consideration: leveraging race conditions, manipulating state, adversary in the middle, authentication bypass, interception, privilege abuse, buffer manipulation, shared resource manipulation, pointer manipulation, excessive allocation, infrastructure manipulation (e.g., cache poisoning), fingerprinting, privilege escalation, protocol manipulation, and obstruction. 

For the same reason, the following hardware meta-patterns may be of interest: leveraging race conditions, manipulating state, privilege abuse, shared resource manipulation, infrastructure manipulation, functionality misuse, privilege escalation, modification during manufacture (a supply chain issue, but worth calling out this specific issue given the potential complexity of detection and limited number of suppliers), hardware integrity attack, obstruction, and hardware fault injection. These two categories have significant overlap given the way in which CAPEC is presented, and this makes sense given our concerns being largely at the hardware/software boundary or below. While primary focus in the CAPEC/CWE framework is on full systems and networks, many of these considerations also apply to sensors -- whether as peripherals or IoT devices.

Communication meta-patterns also overlap with software and hardware meta-patterns. Those worth additional consideration include: exploiting trust in client, adversary in the middle, interception, flooding, excessive allocation, identity spoofing, infrastructure manipulation, footprinting, protocol analysis, communication channel manipulation, protocol manipulation, traffic injection, obstruction, and hardware fault injection. Many of the communications benefits that may come from quantum communications are built atop the concept of quantum entanglement and the inherent protections against interception and tampering it provides. As we've seen, however, these protections are only as strong as the algorithms and protocols used (and their eventual implementations).

These considerations can lead toward a taxonomy such as the initial sketch shown in Table~\ref{tab:table1}. This is not a fully exhaustive taxonomy, but is intended to give an idea of how CWE, CAPEC, and other similar frameworks might be applied to the QIS domains of computation, communications, and sensing. Primarily, these frameworks can be used to provide assurance of comprehensive coverage of potential concerns (as done here), though they could also be used in a more direct manner to drive the structure of the taxonomy.

\begin{table*}[ht]
\centering
\resizebox{\textwidth}{!}{%
\begin{tabular}{llll}\toprule
\textbf{Category} & \textbf{Subcategory}      & \textbf{Vector}                            & \textbf{Impact}                              \\\midrule
Computation & Quadratic Speedup  & Attack on Classical Cryptographic Ciphers \cite{basu_nist_2019} & Unauthorized Decryption                  \\
\cmidrule(lr){2-4}
                  & Maintaining Coherence     & Intentional Interference \cite{cozzolino_highdimensional_2019}              & Denial of Service                            \\
\cmidrule(lr){2-4}
               & Hardware Exploitation    & Flaws (e.g., Spectre, RowHammer) \cite{mutlu_rowhammer_2017}          & Unauthorized Access, Denial of Service \\
\cmidrule(lr){3-4}
                  &                           & Supply Chain Interdiction \cite{zheng_robust_2019}                 & Tampering, Forgery                           \\
\cmidrule(lr){1-4}
Communications  & Quantum Key Distribution  & Fundamental Weakness in Algorithm \cite{patil_comprehensive_2016}         & Unauthorized Decryption                  \\
\cmidrule(lr){3-4}
                  &                           & Exploitation of Flaw in Implementation \cite{facon_detecting_2018}   & Unauthorized Decryption                      \\
\cmidrule(lr){3-4}
                  &                           & Covert Side-channels \cite{curty_foiling_2019} & Eavesdropping \\
\cmidrule(lr){3-4}
                  &                           & Compromise of classical control components \cite{curty_foiling_2019} & Eavesdropping, Tampering, Forgery            \\
\cmidrule(lr){2-4}
                  & Authentication  & Falsified Identity \cite{majumdar_sok_2021}                         & Unauthorized Access                          \\
\cmidrule(lr){2-4}
                  & Interception             & Compromise of Repeater \cite{satoh_attacking_2021}                    & Undetected Eavesdropping, Tampering, Forgery \\
\cmidrule(lr){3-4}
                  &                           & Compromise of Transmission Medium \cite{jogenfors_hacking_2015} & Undetected Eavesdropping                     \\
\cmidrule(lr){1-4}
Sensing  & Accuracy                  & Intentional Interference  \cite{lau_fundamental_2018}          & Denial of Service                            \\
\cmidrule(lr){2-4}
                  & Authenticity \& Integrity & Interdiction of metrology pipeline \cite{shettell_cryptographic_2022}        & Tampering, Forgery                           \\
                  &                           &                                            &    \\\bottomrule                                         
\end{tabular}%
}
\caption{Initial Sketch of a High-level Quantum Security Framework}
\label{tab:table1}
\end{table*}

It would then be possible to consider whether the existing literature leaves threat vectors that are in need of additional attention. Although the specific methodologies to evaluate vectors will depend on their nature and context, general guidance on the expectations for such evaluations should be developed. For example: 

\begin{itemize}
    \item Are computing, communications, and sensing fully representative of the QIS-based systems that should be considered?
    \item When are simulations acceptable versus experiments?
    \item To what extent are mathematical models and proofs expected versus field trials?
    \item What is the metric of success and how is assurance evaluated?
    \item What is the domain of attacker capabilities that are reasonable to consider, and how might that evolve over time?
    \item How should the taxonomy above be kept up-to-date and disseminated?
\end{itemize}

We can then consider existing studies that have attempted to express and quantify the security risks in QIS-based systems. These can be as low-level as the optical channel itself. For example, an attacker might use pulses of light to reveal information about transmitted information such as in a QKD scheme\cite{jain_risk_2015, jogenfors_hacking_2015} or create side-channels for information leakage through intentional damage\cite{makarov_creation_2016}. 

BB84, a provably-secure early quantum encryption protocol, was shown to be susceptible to man-in-the-middle attacks if not properly implemented\cite{huang_man---middle_2009}. Assumptions regarding the implementation of functions such as generating quantum states\cite{xu_experimental_2010} or device calibration\cite{fei_quantum_2018, pljonkin_nonclassical_2021} can also lead to unforeseen potential interception. One must also be careful of making assumptions about the types of attacks an adversary is capable of, lest they not hold in practice\cite{lim_concise_2014}. At least one mathematical framework has been proposed for evaluating the security of quantum channels\cite{ye_generic_2020} but more work is needed in this area. 

Repeaters are a critical component of extending quantum networks beyond point-to-point links of relatively short distances. But they might also present an inherent weak point in the network, at least from an integrity and availability perspective if not confidentiality \cite{satoh_attacking_2021}. Mitigations have been proposed for some weaknesses \cite{satoh_network_2018}, but this is a broad and complex line of inquiry.

Considerations for how the quadratically-faster potential of quantum computers will impact our classical encryption algorithms has been a major point of discussion \cite{vermeer_securing_2020}. The ability of quantum algorithms, namely Shor and Grover, threaten to expose some existing commonly-used algorithms to attack. NIST has led an effort to standardize Post-Quantum Cryptography algorithms, and updated guidance on the usage of classical algorithms\footnote{\url{https://csrc.nist.gov/Projects/post-quantum-cryptography}}.

We must also consider how an attacker could deploy malicious software (malware) to an environment to violate security constraints. For example, interference between qubits could cause ``bit flips'' that could allow observation of a victim process by an attacker-controlled process\cite{sasaki_field_2011}, or interference in a victim process in a manner similar to RowHammer \cite{mutlu_rowhammer_2017}. Automation mechanisms designed to control the complex machinery of a quantum computing device could also be targeted \cite{goncalves_cyberattacks_2021}. This raises the question of whether a different approach (or extensions to existing tools) for ``anti-virus'' or intrusion detection systems (IDS) is needed for quantum information systems \cite{deshpande_towards_2022}.

Similarly, quantum computing promises to provide dramatic performance advantages as it concerns machine learning algorithms \cite{schuld_introduction_2015}. Machine learning algorithms are often built on computationally intensive optimization problems, and thus the speedup of a quantum processor could enable much larger models or data sets. However, classical machine learning is still struggling with a number of challenges regarding explainability\cite{burkart_survey_2021}, input poisoning\cite{jagielski_manipulating_2018}, and privacy\cite{liu_when_2021}. These problems could be carried forward into quantum machine learning algorithms without careful consideration and appropriate countermeasures\cite{kundu_security_2022}.

\section{QIS Cybersecurity Testbed: A Vision}

\begin{figure*}
    \centering
    \includegraphics[width=\textwidth]{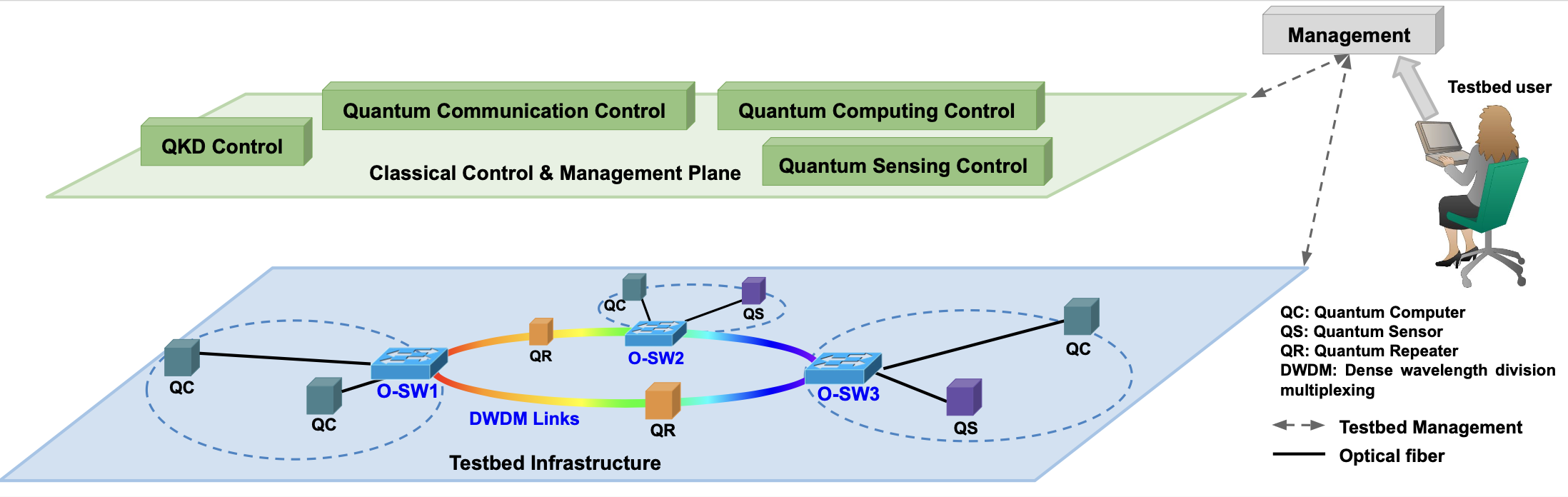}
    \caption{Testbed Design}
    \label{fig:qu-sec-testbed}
\end{figure*}

Existing testbeds tend to have specific purposes and are focused on development of quantum capabilities versus testing security principles. Some of these are simulated to avoid the expense and complexity involved in obtaining, configuring, and maintaining experimental hardware. Simulations, however, are inherently limited in that they can only execute parallel operations to the extent of their parallel processing capabilities\cite{wu_scalable_2022}. True quantum processing and communication protocols depend on superposition states that cannot be fully simulated on non-quantum devices.

Many physical testbed environments are focused on communications. The Secure Communication based on Quantum Cryptography (SECOQC) project in 2004-2008 was a broad coalition of researchers that built a testbed in Vienna\cite{peev_secoqc_2009}. This testbed demonstrated AES-encrypted telecommunications and videoconferencing with keys distributed with a QKD protocol including trusted repeaters. The prototype was limited in its actual transmission capacity and could only provide security guarantees in a controlled network (i.e., no non-trusted nodes). A similar 2011 demonstration in Tokyo was able to demonstrate QKD-secured videoconferencing at a distance of 45~km in a mesh network. The researchers were able to further demonstrate detection of eavesdroppers and rerouting traffic to an uncompromised path\cite{sasaki_field_2011}. In 2014, China Mobile Ltd. demonstrated a QKD-secured network connecting three cities and two metropolitan areas at a total distance of 150~km. The system ran successfully for over 5000 hours\cite{wang_field_2014}. This network resolved several obstacles to broader deployments and resembled a production network in many respects. Taking this further, in 2021 the Madrid Quantum Communication Infrastructure incorporated QKD protocols into a production inter-university network\cite{lopez_madrid_2021}. More recently, a joint effort between QuTech, Eurofiber, and Juniper Networks in the Netherlands have also put together a testbed for a similar purpose \cite{qutech-eurofiber-nl}.

Although many of these studies have been in the context of QKD, the IEQNET testbed (2021) focused on distributing quantum entanglement and demonstrating quantum teleportation, which are relevant to the day-to-day operation of a quantum network. This testbed experimentally showed teleportation over $\sim$44 km with 90\% teleportation fidelity \cite{wu_illinois_2021,ieqnet-arch2022}. A group from Oak Ridge National Laboratory has focused on frequency-related techniques, most recently demonstrating full control of frequency-bin qubits, which is significant to reducing noise in qubit manipulation, and furthering the potential of frequency encoding \cite{PhysRevLett.125.120503}. 
A collaboration between Qubitekk and EPB a developing the first generation of a commercial quantum network in Chattanooga, TN~\cite{earl2022architecture}.
These developments, and others, have led NIST to recognize the increasing maturity of quantum communications and to call for more development of metrology tools and testbeds in which to evaluate them\cite{slattery_quantum_2021}.

Other testbeds are focused on computing capabilities. The Quantum Scientific Computing Open User Tested (QSCOUT) at Sandia National Laboratory is a Department of Energy-funded facility to study the properties of quantum computing devices \cite{clark_engineering_2021}. The immediate goal was to build a quantum computing device with 32 qubits based on trapped ions for experimental purposes. In a similar vein, a number of commercial offerings (e.g., AWS\footnote{\url{https://aws.amazon.com/braket}}, D-Wave\footnote{\url{https://www.dwavesys.com/}},  Google\footnote{\url{https://quantumai.google/}},  Quantinuum\footnote{\url{https://www.quantinuum.com/}}, and IBM \footnote{\url{https://research.ibm.com/quantum-computing}}) have emerged to allow customers to use quantum computing resources, though these are often built as cloud services. 

From the initial taxonomy in Table~\ref{tab:table1} it can be seen that there are many areas where security considerations are relevant. The degree to which these have been experimentally evaluated varies greatly. Many cybersecurity professionals may be interested in learning about these systems and ``red teaming'' against them -- i.e., conducting vulnerability assessments and penetration tests to simulate an attacker and evaluate security controls. The knowledge and resources they have to construct a quantum testbed might, however, be limited due to the high-level of domain-specific knowledge and niche-hardware currently required. There is a need to a ``user-facility'' style testbed where interested researchers can configure test scenarios with systems of interest, run experiments in those scenarios, and collect results to make more formal and quantitative claims about the security strengths or weaknesses of security devices and algorithms. This is similar to the manner in which large-scale super computing resources, particle accelerators, laser sources, and other large and complex open science research infrastructures are operated. 

To meet this need, we suggest a testbed be built and made available to the research community. The diagram in Figure~\ref{fig:qu-sec-testbed} provides a high-level conceptual view of this vision. We would expect such a testbed to include the following elements, for which reason it is more appropriate to think about this as a program than an infrastructure:

\begin{itemize}
    \item Quantum processing equipment - commercial feasibility of these devices currently limits this portion of the testbed to strategic partnerships with organizations that offer quantum computers as a cloud service or have research teams with lab environments open to collaboration. As quantum processors become more commercially viable, incorporating them directly into the testbed itself would be desired.
    \item Quantum sensing - there is an extremely broad array of devices that might be of interest, attempting to procure and maintain a cross-cutting set is not realistic. This is thus another area for strategic partnerships with researchers conducting fundamental science research in areas such as microscopy, gravitation, location-sensing, etc. to pursue a variety of projects assessing the properties of different device types. As this field of devices continues to mature, there may be devices that are representative of a broad class (e.g., using the same underlying chipsets) that would make sense to physically add to the testbed.
    \item Quantum communications - This is the area most suited to building an initial physical testbed due to the availability of commercial optical transmission and sensing components, and Argonne's existing relationships with programs and vendors in this area. For the purposes of evaluating the physical security properties of quantum communications, to include QKD schemes, we would require a pair of single photon detectors (SPD), two polarization analyzers (PA), and an entangled photon source (EPS). Additionally classical computing systems would be required to interact with these components and transmit/receive data, analyze readings from detectors, and attempt to interfere with their operation (which might also require additional EPS, SPD, PA devices or equipment to splice/tap fiber optic cables). 
    \item Layer 1 communications - Quantum repeaters are not yet commercially viable, but are nearing this point. Once they can be obtained they would be a critical evaluation item in a quantum communications testbed. Devices to simulate long-distance fiber-optic cable attenuation, or spools of cable to do so physically, would also be required.
    \item General hardware analysis or reverse engineering equipment - to the extent that physical inspection and validation of hardware components is required (e.g., power analysis, accessing JTAG ports, attempting to cause unexpected behavior more generally), additional lab equipment would be required.
\end{itemize}

This would contribute to the realization of extensible and scalable testing of cybersecurity concerns in quantum computing, communications, and sensors. To support a broad range of experiments, it would support provisioning configurable testbeds at multiple levels, from bare devices for experiments requiring a high degree of control, to functional networks with foundational services for application investigations needing ease of use. To facilitate experiments, it would also provide support services and tools, including software-defined networks, virtualized systems, and real-time monitoring. Last, it would provide quantum device emulation capabilities to capture the (anticipated) behavior of hardware that is not yet available. A quantum cybersecurity testbed resource would allow researchers to develop and explore various QIS components, including hardware devices, communication protocols, network architecture, software stacks, and applications.

\section{Conclusions}

As QIS-based systems continue to mature, it will be important to consider the security implications introduced before they are broadly deployed. Failing to do so may result in adoption of quantum technologies driven by ``hype'' before they are truly secure enough for general use. These products have already entered the commercial space and demonstrated production-grade capabilities. The question would then seem to be not if these technologies become commonplace, but when. In order to establish a solid foundation for comprehensive quantum cybersecurity evaluation that can keep pace with these developments, the taxonomy and testbed introduced in this paper are essential. As in classical information systems, waiting until after adoption can result in a perpetual game of catch-up where security controls are retrofitted and patched into existing systems instead of being incorporated as fundamental components. There is a unique opportunity for quantum information systems to start off on the right foot, but time is of the essence. 

\bibliographystyle{IEEEtran}
\bibliography{qis2022}

\end{document}